\documentclass[twocolumn,showpacs,prb,amsmath]{revtex4-1}
\usepackage{graphicx}
\usepackage{color,amsmath}

\begin{document}

\title{Hot phonon decay in supported and suspended exfoliated graphene}
\author{P. J. Hale$^{*}$, S. M. Hornett, J. Moger, D. W. Horsell, E. Hendry}
\affiliation{School of Physics, University of Exeter, Exeter, EX4 4QL, UK}


\begin{abstract}
Near infrared pump-probe spectroscopy has been used to measure the ultrafast dynamics of photoexcited charge carriers in monolayer and multilayer graphene. We observe two decay processes occurring on 100\,fs and 2\,ps timescales. The first is attributed to the rapid electron--phonon thermalisation in the system. The second timescale is found to be due to the slow decay of hot phonons. Using a simple theoretical model we calculate the hot phonon decay rate and show that it is significantly faster in monolayer flakes than in multilayer ones. In contrast to recent claims, we show that this enhanced decay rate is not due to the coupling to substrate phonons, since we have also seen the same effect in suspended flakes. Possible intrinsic decay mechanisms that could cause such an effect are discussed.
\end{abstract}

\maketitle

The symmetric, linear electronic band structure of graphene gives rise to some very unusual physical properties, such as quantised transmission\cite{Nair2008}, extremely high thermal conductivity\cite{Balandin2008} and high carrier mobility.\cite{Novoselov2004a} An important underlying feature is the very strong electron--phonon coupling that exists in graphene, which is revealed by the presence of Kohn anomalies.\cite{Piscanec2004} In graphite, it is known that strongly coupled optical phonons have high quantum energies of up to 0.2\,eV and are excited only by electrons of elevated energy.\cite{Kampfrath2005,Maultzsch2004,Piscanec2004} To progress towards applications in real (high-current) circuits and devices, it is crucial to understand how graphene behaves under such high energy, non-equilibrium conditions. Despite the surge of interest in this material and its potential applications, investigations into the kinetic properties of ``hot'' charge carriers remain rather limited.\cite{Kampfrath2005,Butscher2007}

Hot electron relaxation in large area, epitaxially grown graphene layers using pump-probe spectroscopy has been studied previously.\cite{Dawlaty2007,George2008,Choi2009,Sun2008,Huang2010,Sun2008,Sun2009,Wang2010,Ruzicka2010,Plochocka2009a} These measurements point to biexponential decay dynamics characterised by a fast $\sim$100\,fs component and a slow $\sim$2\,ps component. There is, however, significant variation in the reported timescales. Epitaxial graphene exhibits inhomogeneity in layer thickness on the micron scale\cite{Huang2010} and can result in significant variability in relaxation dynamics from sample to sample.\cite{Dawlaty2007} Pump-probe measurements have also been performed on mechanically exfoliated graphene,\cite{Newson2009,Zou2010} which is homogeneous over much greater length scales. It was concluded that the slow relaxation process was caused by the coupling to phonons in the substrate.\cite{Newson2009}

In this paper we use near infrared pump-probe spectroscopy to investigate the relaxation dynamics of hot carriers in mechanically exfoliated graphene. Similar to previous results we find that the relaxation occurs on two timescales, one fast ($\sim$100\,fs) and the other slow ($\sim$2\,ps). By measuring the relaxation in monolayer and multilayer graphene flakes we show a clear correlation between the slow decay rate and flake thickness, with the fastest rate observed for monolayer graphene. This slow decay rate is found to occur in both supported and suspended flakes. Therefore, in contrast to recent claims,\cite{Newson2009,Kang2010} the enhanced relaxation is not due to the coupling to substrate phonons.

\begin{figure}[htb]
\includegraphics[width=\columnwidth]{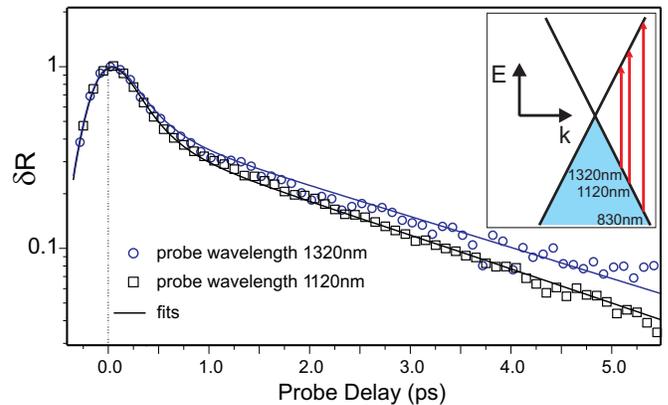}
\caption{(Color online) Differential reflectivity (nomalised at $t=0$) for a 10-layer sample as a function of the probe delay at different probe energies. Inset: the electronic band structure of graphene at the K point of the Brillouin zone with the relevant pump and probe photon energies shown.}
\label{fig1}
\end{figure}

Samples were prepared from mechanically exfoliated natural graphite deposited on 100\,$\mu$m thick glass substrates. The number of layers in each sample was determined by optical contrast\cite{Abergel2007} and four-wave mixing measurements.\cite{Hendry2010} Raman spectroscopy was used to confirm the thickness of monolayers and bilayers from the profile of the 2D peak and the ratio of the G and 2D peak intensities.\cite{Ferrari2006a} Suspended samples were made by depositing flakes on glass substrates that had an array of holes etched into the surface. Square holes between 1.5 and 4.5\,$\mu$m were fabricated by reactive ion etching to a depth of 290\,nm.

To perform the pump-probe measurements, a Ti:sapphire mode-locked laser, center wavelength 830\,nm, pulse width $\sim$180\,fs, was used to optically pump the samples. A probe pulse, $\sim$240\,fs, was generated using an Optical Parametric Oscillator with variable central wavelengths spanning 1100\,--\,1400\,nm. The pump and probe pulses are set with an intensity ratio of $>$\,10:1. The beams were aligned into a microscope and raster scanned through a 1.2\,NA water immersion lens onto the sample, ensuring a spot size $\leq1.5\,\mu$m, which is significantly smaller than the graphene flakes. We estimate the intensity of the combined beams at the sample to be $\sim 0.03\,\text{J/m}^2$. Transient dynamics of the photoelectrons are obtained by measuring the reflected probe light, focussed through 1100\,nm long-pass filters onto a large area photodiode, and recorded as a function of delay between pump and probe signals. The change in the probe light reflectivity is observed to be linearly dependent upon the pump intensity. By varying the focus of the microscope objective lens, we find that the dynamics measured are independent of the size of the excitation region, indicating that lateral charge diffusion does not affect our results.

Figure~\ref{fig1} shows a typical pump-probe measurement of (normalised) differential reflectivity, $\delta R$, for a 10-layer thick sample. When the pump and probe pulses are overlapped in time the pump pulse excites electrons and blocks available transitions for the probe, resulting in a decrease in the reflection from the sample. The relaxation of excited carriers is then observed as a function of time by delaying the probe pulse with respect to the pump and measuring $\delta R$. These data can be described by two exponentially decaying curves, one with a characteristic rate much faster than the other. The fast timescale is not fully resolved in our measurement as it is limited by the probe width. Our interest, however, lies exclusively in the slow timescale which is easily resolved.

\begin{figure}[htb]
\includegraphics[width=\columnwidth]{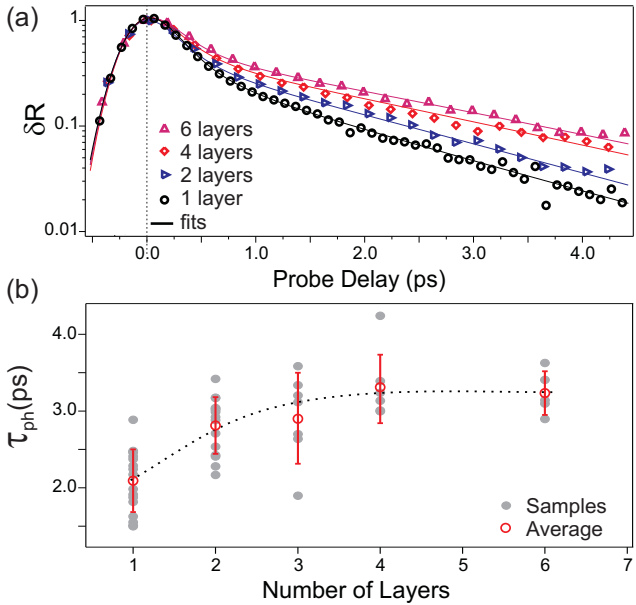}
\caption{(Color online) (a) Differential reflectivity as a function of the probe delay for samples of different thickness, measured using a probe wavelength of 1120\,nm. (b) Extracted phonon lifetimes as a function of layer number. Averaged values per layer are also plotted with the standard deviation as an error bar. The dotted line is a guide to the eye.}
\label{fig2}
\end{figure}

We have performed measurements as a function of the number of graphene layers in a large number of different flakes. Figure~\ref{fig2}(a) contrasts the reflectivity measured in flakes from single to six layers supported on a glass substrate. We observe a clear correlation between layer number and decay rate: faster relaxation occurs for the thinner samples. However, for $\sim$4 layers and above we find that all data collapse onto a single curve. This indicates that at these thicknesses the sample behaves as a bulk material and the rate of relaxation is constant. This suggests that the mechanism that removes energy from the system in thin layers is not available to the thicker layers.

The fast and slow relaxation timescales have been previously attributed to electron--electron and electron--phonon interactions, respectively.\cite{Dawlaty2007} However, it is well known that electron--phonon coupling in graphene features Kohn anomalies\cite{Piscanec2004} that lead to very fast thermalisation with the phonon bath.\cite{Kampfrath2005} Indeed, recent calculations\cite{Butscher2007,Breusing2009} indicate that thermalisation with phonons is completed within the first $\sim$100\,fs following photoexcitation. This suggests that the longer timescale, of the order picoseconds, is associated with the relaxation of the thermalised electron--phonon bath through the decay of hot phonons. 

In order to better understand the mechanisms of the decay, we compare our data to a simple coupled rate equation model which estimates the electron and optical phonon temperatures after photoexcitation from ultrafast optical and near-infrared pulses. This model is based on earlier ones developed in Refs.~\onlinecite{Wang2010} and \onlinecite{Rana2009}, which we outline briefly below. We calculate the phonon generation rate using the relation
\begin{equation} \label{eqn:1}
\Gamma_{\text{ph}}=\alpha \int_{-\infty}^{\infty}E(E-\hbar\omega_\text{ph})\times\,
\left[\rho_{\text{e}} - \rho_{\text{a}}\right]\text{d}E\, ,
\end{equation}
where $\omega_\text{ph}$ is the optical phonon frequency ($\sim$180\,meV), $\rho_{\text{e}}=f(E)(1-f(E-\hbar\omega_\text{ph}))(n_{\text{ph}}+1)$ is the probability of emitting a phonon, $\rho_{\text{a}}=f(E-\hbar\omega_\text{ph})(1-f(E))n_{\text{ph}}$ is the probability of absorbing a phonon and \textit{f}(E) is the Fermi--Dirac distribution of electrons for a given electron temperature, $T_\text{el}$. Here $\alpha\,=\,9/2\times\beta^{2}\left[\pi\rho\omega_{\text{ph}}\hbar^4 v^4_{\text{F}}\right]^{-1}$, is the electron-phonon coupling strength, where $\rho$ is the density of graphene ($\sim7.6\,\times\,10^{-7}\,\text{kg/m}^{2})$, $v_\text{F}$ is the Fermi velocity and $\beta\,=\,45\,\text{eV/nm}$.\cite{Rana2009} The phonon occupation number, $n_{\text{ph}}$, is related to the electron temperature through
\begin{equation} \label{eqn:2}
\frac{\text{d}T_{el}}{\text{d}t}=-2\frac{\Gamma_\text{ph}\hbar\omega_\text{ph}}{C_\text{el}}\,,
\end{equation}
where $C_\text{el}$ is the electronic heat capacity.\cite{Lui2010} (Note that the factor 2 arises from the presence of both $\Gamma$ and K point phonons. Here we treat these phonons as being equivalent due to their similar frequencies.) The rate of optical phonon generation is governed by the equation
\begin{equation} \label{eqn:3}
\frac{\text{d}n_\text{ph}}{\text{d}t}=\frac{\Gamma_\text{ph}}{M_\text{ph}}-\frac{n_\text{ph}-n_\text{ph}^{0}}{\tau_\text{ph}}\,,
\end{equation}
where $n^{0}_\text{ph}$ is the phonon occupation at room temperature, $T_{\text{0}}$, and $\tau_\text{ph}$ is the phonon decay rate. The second term in Eq.~\ref{eqn:3} describes the anharmonic decay of hot phonons. $M_\text{ph}$ is the number of $\Gamma$ and K point phonons (per unit area) that are able to couple to the hot electrons, which is estimated by considering the maximum and minimum momentum possible for an emitted $\Gamma$ or K point phonon.\cite{Wang2010} This gives $M_\text{ph}=\left(2/4\pi\right)\left[(\sqrt{2}E_\text{max}/\hbar v_\text{F})^2-(\omega_\text{ph}/v_\text{F})^2\right]$, where $E_\text{max}$ is upper energy of the hot electons that are able to emit phonons, found by calculating $\Gamma_\text{ph}$ for a given $T_\text{el}$. (Note that the factor of $\sqrt{2}$ in front of $E_\text{max}$ is due to the conservation of pseudospin which limits scattering to states within the same semi-cone in the Brillouin zone, and the factor 2 arises from valley degeneracy.)

Using Eq.~\ref{eqn:2} we can evaluate the temperature of the electron bath as a function of time after excitation. Since the energy from our excitation pulse is dissipated through intraband optical transitions, an initial electron temperature $T_{\text{el}}\text{(0)}$ is calculated from the absorbed fluence and $C_\text{el}$. We account for cooling during the excitation pulse by estimating an average emission of 5 optical phonons per excited electron during the excitation pulse.\cite{Wang2010} Given these starting conditions, a time dependent reflection change is found using
\begin{equation} \label{eqn:4}
\delta R(t)\,\propto\,\tanh\left(\frac{E_\text{pr}}{4k_{B}T_{0}}\right)-\tanh \left(\frac{E_\text{pr}}{4k_{B}T_\text{el}(t)}\right)\, ,
\end{equation}
which gives the change in the strength of interband transitions of the probe pulse due to the electron temperature and the probe photon energy, $E_\text{pr}$. We use probe energies ranging from $0.94-1.11$\,eV ($1120-1320$\,nm). After convolving Eq.~\ref{eqn:4} with the correlation of our pulses, and using just the absorbed pump fluence and $\tau_\text{ph}$ as fitting parameters, we are able to fully describe the time dependent change in reflection measured in our experiments.

The two data sets shown in Fig.~\ref{fig1} are measured with two different probe energies on the same flake: 0.94\,eV and 1.11\,eV. The fast initial decay is roughly the same for both energies ($\leq$\,200\,fs), whereas the slower decay depends on the probe energy. The fits (solid lines) give $\tau_\text{ph}$=3.35\,$\pm$0.1\,ps and $\tau_\text{ph}$=3.6\,$\pm$0.05\,ps for probe energies of 0.94\,eV and 1.11\,eV, respectively. This slow relaxation corresponds to loss of energy from the hot phonon population, presumably through anharmonic decay. This creates a bottleneck for energy loss. Without this finite decay mechanism, hot phonons would accumulate to high occupation numbers, thus changing the timescales of electronic kinetic processes and leading to low damage thresholds in, for example, high current graphene devices.

One can clearly see different dynamics measured for different probe energies. This behaviour arises within our model due to the different electron energies probed in each measurement: higher probe energies, probing higher energy electrons, will yield faster decay dynamics. Our model, meanwhile, yields similar values for the phonon decay time. Theoretical calculations of the hot optical phonon lifetimes in graphene due to anharmonic decay into acoustic phonons were carried out in Ref.~\onlinecite{Bonini2007}, and values between 2 and 4\,ps were reported for both zone-center $E_{\text{2g}}$ and zone-edge $A_{\text{1}}$ modes for phonon temperatures in the $500-900$\,K range. These calculations are clearly consistent with our measurements. Moreover, the values we obtain are similar to those obtained from recent pump-probe measurements on epitaxially grown graphene using a similar model.\cite{Wang2010}

The change in probe reflection as a function of pump delay for our supported flakes of different thickness is shown in Fig.~\ref{fig2}(a). Similarly to Ref.~\onlinecite{Kang2010}, the extracted values of $\tau_\text{ph}$ from fits to the above theory (solid lines) increase as a function of increasing layer number, Fig.~\ref{fig2}(b). The authors suggested that this is a result of the coupling between charge carriers in graphene and phonons in the underlying substrate. In order to investigate the role of the substrate in this effect, the next logical step is to compare the relaxation behaviour observed in supported to that in suspended graphene flakes. 

\begin{figure}[htb]
\includegraphics[width=\columnwidth]{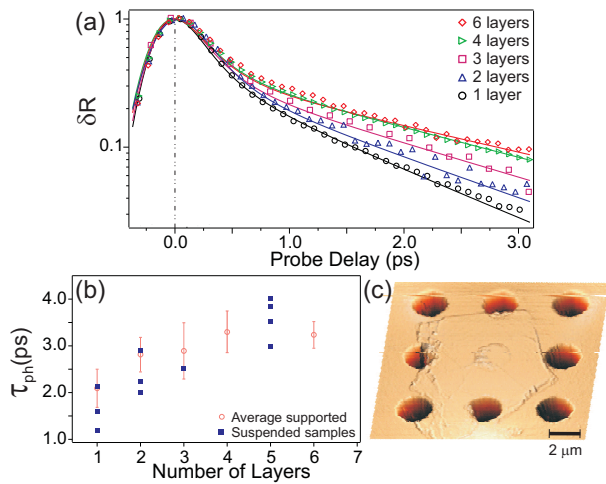}
\caption{(Color online) (a) Differential reflectivity for suspended mono- and few-layer graphene. (b) Extracted phonon lifetimes for mono- and few-layer suspended graphene plotted alongside the averaged values extracted from the supported flakes. (c) An AFM image of a suspended flake covering a 2.5\,$\mu$m, 290\,nm deep hole.}
\label{fig3}
\end{figure}

Figure~\ref{fig3}(a,b) show results obtained from measurements on flakes suspended over 3.5\,$\mu$m holes. Flakes were characterised by atomic force microscopy to establish that they were truly suspended over the holes in the substrates, Fig.~\ref{fig3}(c). The interaction with the substrate phonons will decay on a lengthscale comparable to the phonon wavelength. The maximum wavelength possible for an emitted substrate phonon due to coupling to electrons in graphene is $(2\pi v_{\text{F}}/\omega_{\text{ph}}')$, where $\omega_{\text{ph}}'$ is the substrate optical phonon frequency. For normal optical phonon energies ($\sim$100\,meV) the interaction typically decays over distances $<$ 100\,nm. Since the depth of the etched holes in our case is 290\,nm, we expect interactions with substrate phonons to be substantially suppressed as a relaxation route. In our pump-probe measurements, the high spatial resolution ($\leq$\,1.5\,$\mu$m) also guarantees that only the suspended portions of the flakes are investigated.

Time resolved measurements of the suspended flakes are shown in Fig.~\ref{fig3}(a). Flakes of different thicknesses were measured under the same experimental conditions as the supported flakes. We clearly observe a similar layer dependence as observed for supported layers, with faster decay times observed for thinner flakes. The extracted phonon lifetimes are also similar to those of the supported flakes. In Fig.~\ref{fig3}(b) we compare the suspended measurements to the average value of the supported flakes (with the standard deviation as error bars). Indeed, it would appear that relaxation of suspended monolayer graphene may be even faster than that for the supported flakes. This strongly suggests that the substrate is not responsible for the fast decay observed in thin flakes, and that the mechanism of phonon relaxation in graphene has an intrinsic origin. 

A possible explanation lies with out-of-plane (flexural) phonons, which are suppressed in thicker layers. One would expect flexural modes to be present for monolayer graphene supported on a rough glass substrate as the phonon wavelengths in question are comparable to the roughness of the substrate. Out-of-plane modes are suppressed in thicker layers due to interactions between the graphene planes. As the number of layers increases one would expect an asymptote in the observed dynamics as the system begins to resemble a rigid body. From our data, we suggest that this asymptote is already reached at approximately four layers.

In conclusion, near infrared pump-probe spectroscopy has been used to measure the ultrafast dynamics of photoexcited charge carriers in monolayer and multilayer graphene. We observe two timescales in the decay occurring on $\sim$100\,fs and $\sim 2-3$\,ps. The fast relaxation timecale can be attributed to electron--phonon thermalization, whereas the slower timescale represents a bottleneck in the relaxation process due to anharmonic decay of hot phonons. Using a simple theoretical model we have calculated the decay rates in different graphene flakes and shown that the hot phonon decay is faster in monolayer than in multilayer graphene. Comparing our results on supported and suspended flakes we have demonstrated that substrate phonons are not the mechanism for removing energy from the system. The possible origin of the intrinsic mechanism is the enhanced coupling to out-of-plane flexural phonon modes.

This work was funded by the EPSRC (Grant No. EP/G036101/1 and EP/G041482/1) and RCUK. The authors would like to thank V. Falko, E. Mariani, M. Portnoi and A. Shytov for useful discussions. PJH would also like to thank S. Landi and T. Khodkov for assistance with fabrication. Special thanks to A. K. Savchenko, who will be sadly missed.\\
\\
\noindent$^{*}$P.J.Hale@ex.ac.uk

\end{document}